\documentclass[reprint,amssymb,amsmath,aps,showpacs,10pt,floatfix,longbibliography,pra,floatfix,nofootinbib]{revtex4-2}

\usepackage{graphicx}
\usepackage{color}
\usepackage[colorlinks,bookmarks=false,citecolor=darkblue,linkcolor=red,urlcolor=blue]{hyperref}

\definecolor{darkblue}{rgb}{0,0.02,0.45}

\graphicspath{{../figures/}}

\begin{document}

\title{Kitaev magnetism through the prism of lithium iridate}

\author{Alexander A. Tsirlin}
\email{altsirlin@gmail.com}

\author{Philipp Gegenwart}
\affiliation{Experimental Physics VI, Center for Electronic Correlations and Magnetism, University of Augsburg, 86159 Augsburg, Germany}

\begin{abstract}
Fifteen years since its inception, Kitaev model still boasts only a narrow group of material realizations. We review the progress in studying and understanding one of them, lithium iridate Li$_2$IrO$_3$ available in three polymorphs that host strong Kitaev interactions on spin lattices of different dimensionality and topology. We also discuss feasibility, effectiveness, and repercussions of tuning strategies based on the application of external pressure and chemical substitutions.
\end{abstract}

\maketitle

\section{Introduction}

Quantum states characterized by long-range entanglement show some of the most intriguing and enigmatic properties in condensed-matter systems~\cite{wen2019}. Understanding the nature of these quantum states and elucidating their behavior remains a formidable challenge, because relevant microscopic models are notoriously difficult or even impossible to solve. In this context, Alexei Kitaev's discovery of an analytically solvable spin model hosting a quantum spin liquid ground state received significant attention, reflected for example in 2300 citations of his original work~\cite{kitaev2006} over the last 15 years and in Kitaev interactions emerging as the new type of magnetic exchange on par with the Heisenberg and Dzyaloshinskii-Moriya interactions as textbook examples of magnetic couplings.

Kitaev spin model was originally formulated for the planar honeycomb lattice and later extended to different tricoordinated lattices in three dimensions~\cite{mandal2009,obrien2016}. The crux of the model is the Ising-like exchange term where spin direction $\gamma$ varies from one bond to another,
\begin{equation}
 \mathcal H=K\sum_{\langle ij\rangle} S_i^{\gamma}S_j^{\gamma}
\label{eq:kitaev}
\end{equation}
($\gamma=X,Y,Z$ depending on the bond). This special setting creates an exchange frustration, which is quite different from the more common geometrical frustration, and implies that long-range magnetic order does not form because no stable spin direction can be found in the magnetically ordered state. On a more rigorous level, the solution of Eq.~\eqref{eq:kitaev} is facilitated by representing each spin operator as a combination of four Majorana fermions that are separated into flux variables and free ``matter'' fermions. This elegant construction has far-reaching implications. Excitations of the Kitaev spin liquid exhibit an unusual anyonic statistics (they are neither bosons nor fermions) and create interesting opportunities for topological quantum computing~\cite{kitaev2006}.

Different extensions of the Kitaev model were considered over the last years. Details of these extended models, along with basic properties of the parent Kitaev model itself, can be found in several review articles~\cite{hermanns2018,janssen2019,motome2020}. Another interesting direction has been the search for real-world manifestations of the Kitaev physics and, specifically, for insulating magnets dominated by the Kitaev exchange term. 

Bringing Kitaev model into the lab requires a spin-orbit-entangled electronic state of the transition-metal ion. The symmetries of this state and associated exchange pathways should favor the interaction term of Eq.~\eqref{eq:kitaev} and prevent Heisenberg or any other type of exchange. This situation is rather unique and was thought to be accomplishable only in $4d$ and $5d$ metals with the low-spin $d^5$ state and strong spin-orbit coupling~\cite{jackeli2009}. More recently, the possibility of a similar microscopic scenario in the high-spin $d^7$ states~\cite{liu2018} and selected $4f$ states~\cite{jang2019,motome2020b} was put forward, but at the time of this review most experimental efforts are still focused on a handful of $d^5$ compounds of Ir$^{4+}$ ($5d$) and Ru$^{3+}$ ($4d$). The purpose of the present article is to look at some interesting aspects and associated challenges through the prism of a single material, lithium iridate Li$_2$IrO$_3$, that adopts three different crystal structures, allows multiple chemical substitution, and highlights several generic features of Kitaev magnets. For a recent review of other Kitaev materials see also Refs.~\cite{winter2017,takagi2019}.

\section{Parent compounds}

\subsection{Crystal structure}
Two decades ago solid-state chemists would probably find lithium iridate a very mundane compound, yet another derivative of the rocksalt structure with alkaline and transition metals filling octahedral voids of the close-packed oxygen framework~\cite{hauck1980,mather2000}. Similar compounds are known for many of the $4d$, $5d$, and even $5p$ metals~\cite{delmas1980} and show little differences in their crystal chemistry. This fact, along with the high price of iridium and its undesirability for material science, did not motivate detailed studies. Indeed, no single attempt of the crystal structure refinement for Li$_2$IrO$_3$ was performed until 2000's~\cite{kobayashi2003,malley2008}, despite the first reports on the synthesis having appeared much earlier~\cite{gadzhiev1984,kobayashi1997}.

\begin{figure*}
\centerline{\includegraphics{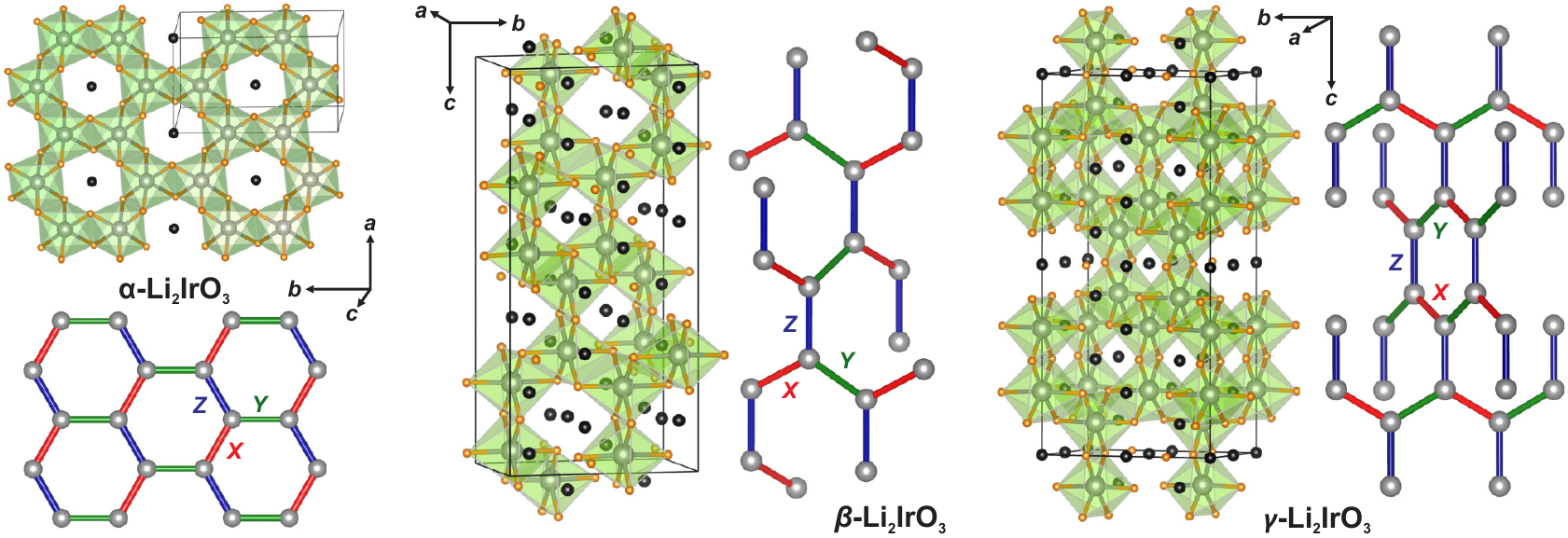}}
\caption{\label{fig:structure}
Crystal structures of three Li$_2$IrO$_3$ polymorphs and the respective magnetic topologies: planar honeycomb ($\alpha$), hyperhoneycomb ($\beta$), and stripy-honeycomb ($\gamma$) lattices of the Ir$^{4+}$ ions. Different colors show $X$-, $Y$-, and $Z$-bonds of the Kitaev model, Eq.~\eqref{eq:kitaev}.
}
\end{figure*}

Recent interest triggered by prospects of the Kitaev physics has led to an extensive reinvestigation of the layered $\alpha$-Li$_2$IrO$_3$ polymorph that was shown to be magnetically ordered below $T_N\simeq 15$\,K~\cite{singh2012}, in contrast to the earlier report where no magnetic order was found~\cite{felner2002}. Further research culminated in the discovery of two new polymorphs~\cite{biffin2014a,takayama2015,modic2014} that differ in the spatial arrangement of Li and Ir within the octahedral voids. This difference leads to distinct crystallographic symmetries and magnetic topologies (Fig.~\ref{fig:structure}). Monoclinic $\alpha$-Li$_2$IrO$_3$ has planar honeycomb structure in the vein of the original Kitaev model. In contrast, orthorhombic $\beta$- and $\gamma$-Li$_2$IrO$_3$ feature three-dimensional (3D) hyperhoneycomb and stripy-honeycomb (or harmonic-honeycomb) lattices, respectively. Theoretical work on 3D versions of the Kitaev model surged~\cite{nasu2014a,nasu2014b,lee2014a,lee2014b,lee2015,perreault2015,kimchi2015} after these new polymorphs were discovered. 

Polymorphism related to the different distribution of cations within the closed-packed oxygen framework is not uncommon for ternary oxides. Some of them even adopt same symmetries as $\alpha$- and $\beta$-Li$_2$IrO$_3$~\cite{mather2000}, although detailed knowledge on thermodynamic stability of different polymorphs in relation to their chemistry and composition is presently missing. In the Li$_2$IrO$_3$ case, $\alpha$- and $\gamma$-polymorphs transform into the $\beta$-polymorph upon heating above 1000\,$^{\circ}$C~\cite{biffin2014a,freund2016,ruiz2020}, suggesting that $\beta$-Li$_2$IrO$_3$ should be the thermodynamically stable form of lithium iridate.

\begin{figure}
\centerline{\includegraphics{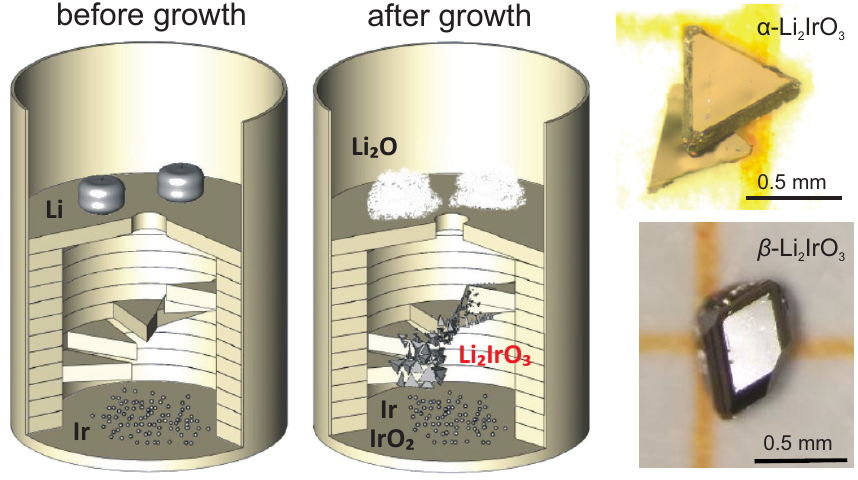}}
\caption{\label{fig:growth}
Crystal growth of $\alpha$- and $\beta$-Li$_2$IrO$_3$ from separated educts. Metallic Li and Ir are placed in different parts of the crucible with intermediate spikes serving as crystallization centers. The typical crystals are shown on the right. Part of the figure is adapted from Ref.~\cite{freund2016} (CC-BY).
}
\end{figure}

In recent years, Li$_2$IrO$_3$ gained its appeal also among solid-state chemists. New experiments on Li deintercalation~\cite{mccalla2015,pearce2017,hong2019} highlighted the unusual anion-redox mechanism, the partial oxidation of oxygen upon extraction of Li, that can be used to increase reversible capacity of battery materials~\cite{grimaud2016}. These two lines of researchs, batteries and magnetism, developed entirely independently from each other and have yet to explore possible interesting connections, such as tuning or charge doping of Li$_2$IrO$_3$ via controllable electrochemical intercalation. 


\subsection{Crystal growth}

Iridate compounds are notorious for their difficult crystal growth, but Li$_2$IrO$_3$ is exceptional even among those. The proclivity of Ir$^{4+}$ toward reduction and oxygen release prevents the usage of floating-zone method, because oxygen pressures required to stabilize the Ir$^{4+}$ state are very high and essentially unfeasible. In the case of lithium iridate, decomposition takes place around 1200\,$^{\circ}$C well before melting point is reached. On the other hand, sizable vapor pressures of both alkaline-metal oxides and iridium oxide below decomposition temperature facilitate crystal growth from gas phase via some kind of vapor transport. This process is most prominent in Na$_2$IrO$_3$ where mm-size single crystals are grown by a simple annealing of pre-reacted powder sample~\cite{singh2010}.

The growth of Li$_2$IrO$_3$ is more tricky in the sense that only tiny crystals with the lateral dimensions of 0.1-0.2\,mm can be obtained by simple annealing~\cite{ruiz2017,ruiz2020}. Moreover, crystals of the $\alpha$-polymorph can not be grown in this way, because vapor transport remains weak in the temperature range where this phase is stable. The method of separated educts~\cite{freund2016} leads to improved results and increases the typical crystal size to about 0.5\,mm. Here, lithium and iridium metals are used as reactants and separated in space inside the crucible. Spikes are deliberately placed between the reactants and serve as crystallization centers during the vapor growth from oxidized Li and Ir (Fig.~\ref{fig:growth}). Changing reaction temperature allows growth of both $\alpha$- and $\beta$-polymorphs~\cite{freund2016,majumder2019,freund}.

Whereas $\beta$- and $\gamma$-Li$_2$IrO$_3$ are believed to be free from crystal defects, $\alpha$-Li$_2$IrO$_3$ crystals suffer from both twinning and stacking faults~\cite{freund2016,freund}. While the latter is a notorious problem of all layered materials, the former results from an accidental match between the $a$ and $c$ lattice parameters. This twinning issue is unique to $\alpha$-Li$_2$IrO$_3$ and renders it a less convenient Kitaev material, especially in the context of various tuning attempts discussed in Sec.~\ref{sec:tuning} below. Crystal quality has drastic influence on the magnetic behavior. Stacking faults broaden magnetic ordering transition and eventually eliminate it~\cite{manni}, resulting in a state that ostensibly resembles a spin liquid but hardly constitutes one.


\subsection{Magnetic anisotropy}

\begin{figure*}
\centerline{\includegraphics{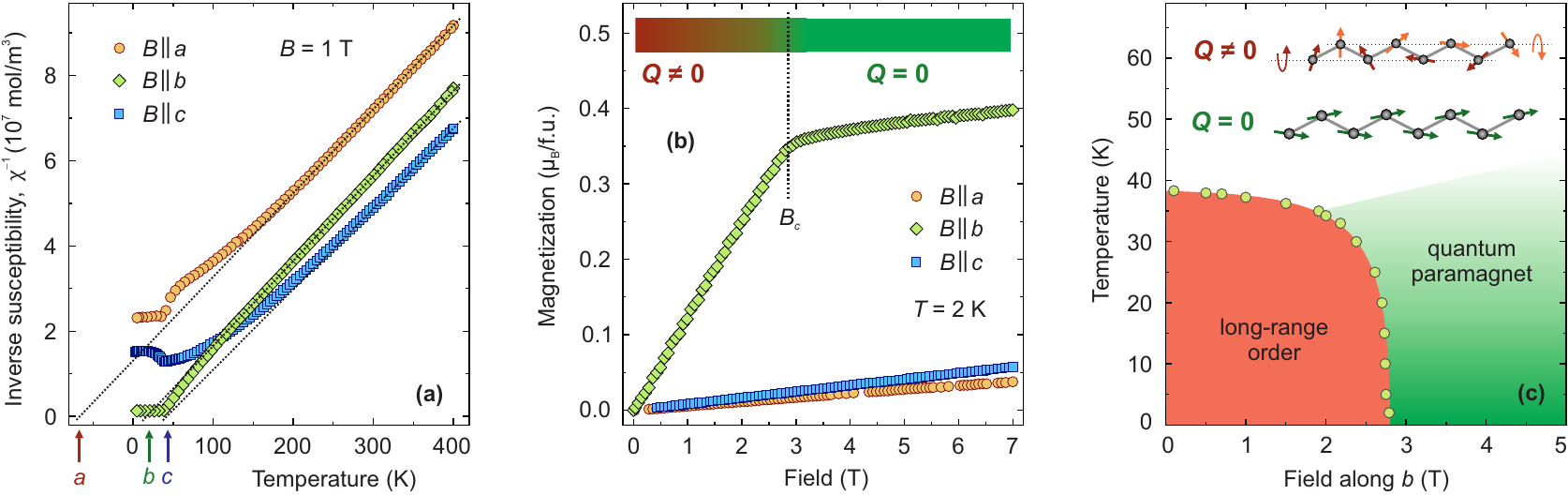}}
\caption{\label{fig:properties}
Anisotropic magnetic response of $\beta$-Li$_2$IrO$_3$~\cite{majumder2019}. (a) Inverse magnetic susceptibility for three field directions; the Curie-Weiss fits (dotted lines) have a similar slope but very different intercepts (Curie-Weiss temperatures). (b) Field-dependent magnetization reveals the transition at $B_c\simeq 2.7$\,T for $B\,\|\,b$; the color bar illustrates a gradual replacement of the incommensurate $Q\neq 0$ order by the commensurate $Q=0$ mode of the quantum paramagnet that does not break any symmetries and, therefore, does not cause a magnetic transition as a function of temperature above $B_c$. (c) Temperature-field phase diagram for $B\,\|\,b$; the phase boundary is probed by thermodynamic measurements. The $Q\neq 0$ (counter-rotating spirals) and $Q=0$ modes are shown schematically and do not reflect all details of the real magnetic structure~\cite{biffin2014a}. 
}
\end{figure*}

Magnetic response of the Li$_2$IrO$_3$ polymorphs is highly anisotropic (cf. Fig.~\ref{fig:properties} for the $\beta$-polymorph). At high temperatures, paramagnetic effective moments are in the range of $1.5-2.0\,\mu_B$ for all field directions~\cite{singh2012,mehlawat2017,majumder2019,modic2014}, as expected for Ir$^{4+}$ in the regular octahedral environment ($j_{\rm eff}=\frac12$ state independently confirmed by resonant inelastic x-ray spectroscopy~\cite{gretarsson2013,takayama2019} and optical measurements~\cite{hinton2015,li2017,li2015,hermann2017}). This renders on-site magnetic anisotropy, such as $g$-tensor anisotropy, weak. A much stronger anisotropy arises from the exchange couplings. Curie-Weiss temperatures for the same polymorph not only differ in absolute values, but also show different signs depending on the field direction (Fig.~\ref{fig:properties}a)~\cite{majumder2019,modic2014}. While this anisotropy potentially carries useful information on the size of Kitaev and other exchange terms~\cite{winter2016}, their direct calculation from the magnetic susceptibility proved to be difficult. The problem may lie in single-ion effects (electronic excitations at the Ir$^{4+}$ site) that cause deviations from the simple Curie-Weiss behavior~\cite{li2021}. 

$\beta$- and $\gamma$-polymorphs additionally show a very anisotropic response in field-dependent magnetization at low temperatures~\cite{modic2014,ruiz2017}. The field applied along one special direction, which is crystallographic $b$ for the crystal structures shown in Fig.~\ref{fig:structure}, leads to a partial polarization with a field-induced transition around $B_c=2.7$\,T (Fig.~\ref{fig:properties}b) accompanied by a seeming loss of magnetic order: no phase transition as a function of temperature can be seen for this field direction above $B_c$~\cite{ruiz2017,modic2017,majumder2019}. This behavior is often taken as a hallmark of Kitaev physics, because several candidate materials, including $\alpha$-RuCl$_3$~\cite{wolter2017,sears2017,bachus2020} and, more recently, Co$^{2+}$ oxides~\cite{zhong2020,yao2020}, all revealed a similar evolution. Naively, one may expect Kitaev spin liquid to appear in $\beta$-Li$_2$IrO$_3$ above $B_c$ for $B\,\|\,b$, but this field-induced state is in fact a rather mundane quantum paramagnet, as we will see shortly. 

Another important remark at this point is that magnetic anisotropy, including anisotropy of the magnetization, is not a consequence of strong Kitaev interactions \textit{per se}. Despite the underlying exchange anisotropy, Kitaev model itself [Eq.~\eqref{eq:kitaev}] is in fact isotropic and would show equal Curie-Weiss temperatures for all directions of the applied field~\cite{winter2016}. The strongly anisotropic magnetic response reviewed in this section indicates salient deviations of all Li$_2$IrO$_3$ polymorphs from the simple Kitaev model.


\subsection{Magnetic order and interactions}
\label{sec:order}

The departure of all Li$_2$IrO$_3$ polymorphs from the pure Kitaev model is cemented by the presence of long-range magnetic order below $T_N=15$\,K~\cite{williams2016} ($\alpha$), $37-38$\,K~\cite{biffin2014a} ($\beta$), and 40\,K~\cite{biffin2014b} ($\gamma$). Magnetic structures are incommensurate, non-coplanar, and bear surprising resemblances to each other. All of them feature zigzag chains with counter-rotating spin spirals (Fig.~\ref{fig:properties}c), quite unusual magnetic entities that signal strong Kitaev interactions and should not appear in conventional antiferromagnets dominated by Heisenberg exchange~\cite{kimchi2015}. 

Further experiments showed that the field-induced state of $\beta$-Li$_2$IrO$_3$ for $B\,\|\,b$ and $B>B_c$ lacks long-range magnetic order but features strong commensurate ($Q=0$) correlations, with ferromagnetic alignment of the $b$-components of the spin and antiferromagnetic, zigzag-like alignment of the $ac$-components~\cite{ruiz2017}. When magnetic field is applied along the $b$ direction, such a state does not break any symmetries and does not lead to a phase transition upon cooling (Fig.~\ref{fig:properties}c). It is in fact similar to a fully polarized state of an isotropic ferromagnet. This explains the apparent absence of magnetic order above $B_c$ in the sense that no phase transition takes place as a function of temperature; phase transition is replaced by a crossover. The resulting low-temperature state is by no means a spin liquid and should be classified as quantum paramagnet~\cite{majumder2019}.

The type of magnetic order can be indicative of underlying magnetic interactions. In the case of $j_{\rm eff}=\frac12$ moments of Ir$^{4+}$, at least three main terms have to be considered~\cite{rau2014},
\begin{equation}
 \mathcal H_{12}=J\mathbf S_1\mathbf S_2+KS_1^{\gamma}S_2^{\gamma}+\Gamma(S_1^{\alpha}S_2^{\beta}+S_1^{\beta}S_2^{\alpha}),
\label{eq:jkg}\end{equation}
where $J$ is Heisenberg exchange, $K$ is Kitaev exchange, and $\Gamma$ is off-diagonal anisotropy. The index $\gamma$ is one of the Kitaev axes ($X,Y,Z$) and changes from bond to bond, while $\alpha$ and $\beta$ denote spin components perpendicular to $S^\gamma$. 

Magnetic order observed in $\beta$- and $\gamma$-Li$_2$IrO$_3$ can be rationalized within the framework of this minimal $JK\Gamma$ model for nearest-neighbor interactions and with ferromagnetic Kitaev term $K$~\cite{lee2016,ducatman2018,stavropoulos2018}. This model also explains similarities of magnetic order across the different polymorphs~\cite{krueger2020}. Moreover, field-induced phase transition and the $B_c$ value of $\beta$-Li$_2$IrO$_3$ gauge $J\simeq 0.3$\,meV~\cite{rousochatzakis2018}, suggesting that at least $\beta$-Li$_2$IrO$_3$ should be dominated by $K$ and $\Gamma$, while $J$ is negligible. The smallness of an excitation gap below $B_c$~\cite{majumder2020} puts an additional $|K|\simeq |\Gamma|$ constraint and allows a quite accurate positioning of $\beta$-Li$_2$IrO$_3$ on the phase diagram, in reasonable agreement with the \textit{ab initio} results~\cite{katukuri2016,majumder2018}. The absolute values of both $|K|$ and $|\Gamma|$ are in the range of $10-15$\,meV~\cite{majumder2018,majumder2019}.

Less is known about the microscopic parameters of $\alpha$-Li$_2$IrO$_3$. Magnetic structure of this compound was so far rationalized only within an extended model that breaks equivalence of interactions on the $XY$- and $Z$-bonds of the honeycomb lattice~\cite{williams2016}. Such disparity is indeed allowed by symmetry in all Li$_2$IrO$_3$ polymorphs (and also in other Kitaev candidates), but its role is generally believed to be minor~\cite{winter2016}. The situation may be different in $\alpha$-Li$_2$IrO$_3$, though. Moreover, the terms beyond Eq.~\eqref{eq:jkg}, including Dzyaloshinskii-Moriya interactions and long-range Heisenberg interactions, may not be insignificant~\cite{winter2016}. It is also worth noting that a large deal of confusion in the microscopic analysis~\cite{winter2016,nishimoto2016} was caused by the inaccurate crystallographic data used for $\alpha$-Li$_2$IrO$_3$ prior to 2016. The results obtained for crystal structures other than the experimental one from Ref.~\cite{freund2016} (or, alternatively, the fully relaxed one~\cite{winter2016}) may be strongly influenced by this ambiguity.


\subsection{Magnetic excitations}
\label{sec:excitations}

Incommensurate long-range order causes complex magnon spectra that were indeed observed in the energy range up to 12\,meV in $\alpha$-Li$_2$IrO$_3$~\cite{choi2019} and up to 16\,meV in $\beta$-Li$_2$IrO$_3$~\cite{ruiz2021}. Despite the strong exchange anisotropy, magnon gaps in zero field are negligible~\cite{choi2019,majumder2020}. In the case of $\beta$-polymorph, field-induced phase transition opens a gap above $B_c$, similar to another Kitaev material $\alpha$-RuCl$_3$ where this gap opening has been studied extensively~\cite{wang2017,sahasrabudhe2020,wulferding2020}. 

The presence of sizable or even dominant Kitaev interactions led to an idea that excitations other than magnons may appear at higher energies above the one-magnon part of the spectrum, or above $T_N$ where long-range magnetic order disappears. Indeed, an excitation continuum has been seen above 20\,meV in different Li$_2$IrO$_3$ polymorphs using resonant inelastic x-ray scattering~\cite{revelli2020,ruiz2021} and Raman spectroscopy~\cite{gupta2016,glamazda2016,li2020a}. This phenomenology resembles not only other Kitaev materials~\cite{banerjee2017,kim2020}, but also frustrated magnets with predominant Heisenberg interactions~\cite{li2020}. Consequently, it remains debated whether the excitation continuum manifests Majorana excitations of the underlying Kitaev model~\cite{nasu2016} or arises from magnon breakdown triggered by the strong exchange anisotropy~\cite{winter2017b}.

Interestingly, the continuum persists well above $T_N$~\cite{revelli2020,ruiz2021,kim2020} where magnon description should become irrelevant. Calculations for the pure Kitaev model reveal interesting similarities to the experimental observations in this temperature range~\cite{revelli2020,kim2020}, although it remains unclear whether additional interactions manifested by the spectral structure below $T_N$~\cite{kim2020} can be simply neglected when magnetic order disappears. Calculating finite-temperature response for extended Kitaev models~\cite{winter2018} remains a significant challenge that essentially constraints possible interpretations of the continuum to fermionic excitations of the pure Kitaev model, whereas other scenarios await a thorough theoretical study. The excitation continuum present far above $T_N$ appears to be a generic feature of the Kitaev iridates and may be used to classify them as ``proximate Kitaev spin liquids'', although details of this purported connection remain to be explored.


\subsection{Open questions}

Finally, we summarize main open questions regarding three Li$_2$IrO$_3$ polymorphs. First, the exact microscopic scenario of $\alpha$-Li$_2$IrO$_3$ is far from being settled. This Kitaev material was among the first to appear~\cite{singh2012} and initially appealed researchers with its apparent simplicity, but nonetheless it eludes complete understanding even after a decade of extensive research. It also shows the peculiar spatial anisotropy of the underlying honeycomb lattice. This effect was not seen in any other Kitaev material and deserves a more thorough investigation, especially in the light of newly prepared substituted compounds (Sec.~\ref{sec:substitutions}) where pre-conditions for deviating from isotropic Kitaev interactions become even stronger.

Second, $\beta$-Li$_2$IrO$_3$ shows one interesting feature that goes beyond the already established microscopic scenario. A very weak remnant magnetization develops below 100\,K, well above $T_N=37-38$\,K. Muon spin relaxation confirms bulk nature of this putative order, although no oscillations typical for a long-range-ordered state appear at 100\,K. An unusual type of order may thus coexist or probably compete with the conventional incommensurate long-range order that develops below $T_N$~\cite{ruiz2020}. Further experiments, such as thermal expansion measurements, may be interesting to probe the effect of this 100\,K anomaly on the lattice. It is also interesting whether aspects like Li deficiency or Li dynamics can be relevant in this case.

Third, the incommensurate non-coplanar order of Li$_2$IrO$_3$ is quite unusual in its own right. The common wisdom is that non-collinear magnetic orders are very sensitive to the applied field and readily turn into collinear spin-density waves~\cite{pregelj2015} or exotic spin textures like skyrmions~\cite{tokura2021}. At first glance, the field-induced transition of $\beta$-Li$_2$IrO$_3$ in $B\,\|\,b$ seems fairly mundane because it leads to a state with commensurate spin-spin correlations, but the way this new state sets in is also quite unusual. Commensurate correlations appear already below $B_c$ and compete with the incommensurate order~\cite{ruiz2017}. This is very different from conventional antiferromagnets where collinear state abruptly transforms into a spin-flop state when field is applied along the magnetic easy axis.

Theoretical studies based on the $JK\Gamma$ model highlight an interesting evolution of incommensurate Li$_2$IrO$_3$ orders in magnetic field~\cite{li2020b}. $\beta$-Li$_2$IrO$_3$ is predicted to undergo a field-induced transition also for $B\,\|\,c$ around 13\,T~\cite{li2020b}, the field feasible for different experimental probes of magnetic order and phase transitions. In the case of $\gamma$-Li$_2$IrO$_3$, torque measurements in high fields revealed an unusual sawtooth shape discussed in the context of chiral spin order~\cite{modic2018}, although calculations for realistic spin models with strong exchange anisotropy offer a more conventional explanation of this finding~\cite{riedl2019}.

\section{Tuning}
\label{sec:tuning}

\begin{figure}
\centerline{\includegraphics{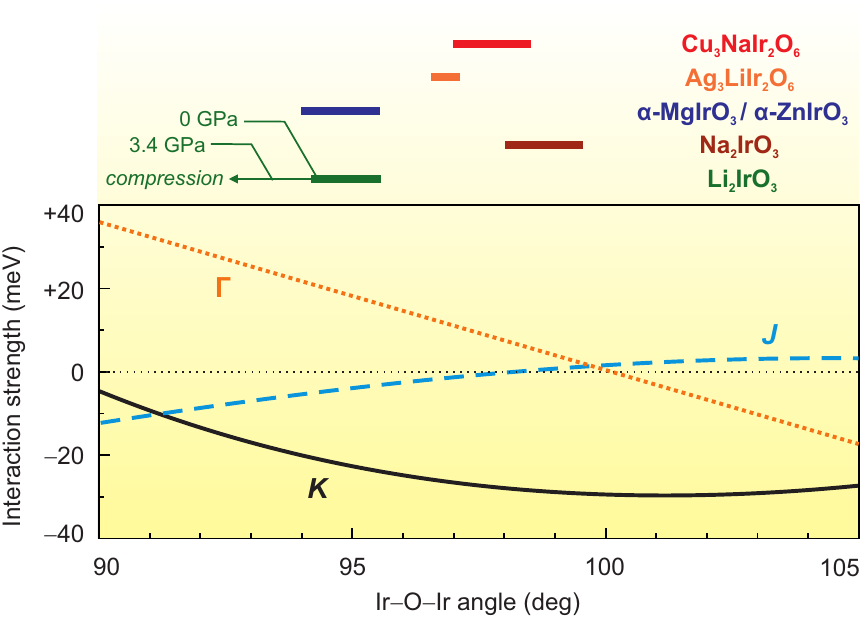}}
\caption{\label{fig:geometry}
Interaction parameters of Eq.~\eqref{eq:jkg} vs. Ir--O--Ir bridging angle~\cite{winter2016}. Kitaev limit is approached around 100\,deg. Colored bars schematically show the bridging angles in some of the candidate materials listed in Table~\ref{tab:compounds}. The bar length reflects the distribution of angles between the $X$-, $Y$-, and $Z$-bonds and between compounds of the same series: $\alpha$-Li$_2$IrO$_3$~\cite{freund2016} and $\beta$-Li$_2$IrO$_3$~\cite{takayama2015}, Na$_2$IrO$_3$~\cite{choi2012}, $\alpha$-MgIrO$_3$ and $\alpha$-ZnIrO$_3$~\cite{haraguchi2018}, Ag$_3$LiIr$_2$O$_6$~\cite{bahrami2019}, and Cu$_3$NaIr$_2$O$_6$~\cite{roudebush2016}. Pressure evolution is shown for $\beta$-Li$_2$IrO$_3$ based on the data from Ref.~\cite{majumder2018}.
}
\end{figure}

The presence of magnetic long-range order, as well as other apparent deviations of all Li$_2$IrO$_3$ polymorphs from the pure Kitaev model, raised the question whether a suitable tuning could enhance Kitaev interaction term $K$ at the cost of reducing the $J$ and $\Gamma$ terms of Eq.~\eqref{eq:jkg}. On the microscopic level, one expects $J$, $K$, and $\Gamma$ to be strongly dependent on the local geometry and especially on the Ir--O--Ir bridging angles in the crystal structure (Fig.~\ref{fig:geometry})~\cite{winter2016,nishimoto2016}. 

The $\alpha$- and $\beta$-polymorphs of Li$_2$IrO$_3$ feature the bridging angles of $94-95^{\circ}$ that should result in $K<0$ and $\Gamma\simeq |K|$ with a relatively smaller $|J|\ll |K|$~\cite{winter2016}. This interaction regime was established for $\beta$-Li$_2$IrO$_3$ indeed and can be expected for the other polymorphs, too (Sec.~\ref{sec:order}). Reducing the bridging angle toward $90^{\circ}$ has the unwanted effect of lowering $K$ and enhancing $\Gamma$. On the other hand, increasing the angle toward $100^{\circ}$ seems to be the right choice. Here, different hopping parameters may conspire to suppress $J$ and $\Gamma$, while keeping $K$ large and ferromagnetic~\cite{rau2014,winter2016}, manifesting proximity to the Kitaev limit.

In the following, we review different tuning strategies from this microscopic perspective. Pressure experiments define how the bridging angles and, consequently, interaction terms of Eq.~\eqref{eq:jkg} are affected by changes in the unit cell volume. Volume changes are then used to assess different compounds prepared by chemical substitution.


\subsection{Pressure and Strain}

Pressure has drastic effect on Li$_2$IrO$_3$. Structural phase transition detected at room temperature around $P_c=3.8$\,GPa in $\alpha$-Li$_2$IrO$_3$ ($C2/m\rightarrow P\bar 1$)~\cite{hermann2018} and $3.8-4.4$\,GPa in $\beta$-Li$_2$IrO$_3$ ($Fddd\rightarrow C2/c$)~\cite{takayama2019,veiga2019,choi2020} leads to magnetism collapse via an abrupt shortening of $\frac13$ of the Ir--Ir bonds on the (hyper)honeycomb lattice. The high-pressure phase is dimerized and entirely non-magnetic, because unpaired electrons of Ir are now involved into the formation of metal-metal bonds manifested by the short Ir--Ir distances in the crystal structure. 

This instability seems to be generic for all Kitaev magnets. It has been observed in $\alpha$-RuCl$_3$ already below 1\,GPa~\cite{bastien2018,biesner2018} and can be expected in Na$_2$IrO$_3$ too, albeit at much higher pressures~\cite{hu2018} where experimental situation remains controversial~\cite{hermann2017,xi2018,layek2020}. Prominent changes in the electronic spectrum measured by resonant inelastic x-ray scattering, especially the suppression of the characteristic peak of the $j_{\rm eff}=\frac32\rightarrow\frac12$ excitation around 0.7\,eV, confirm the breakdown of the $j_{\rm eff}=\frac12$ state of iridium in the dimerized phase~\cite{clancy2018,takayama2019}.

\begin{figure}
\centerline{\includegraphics{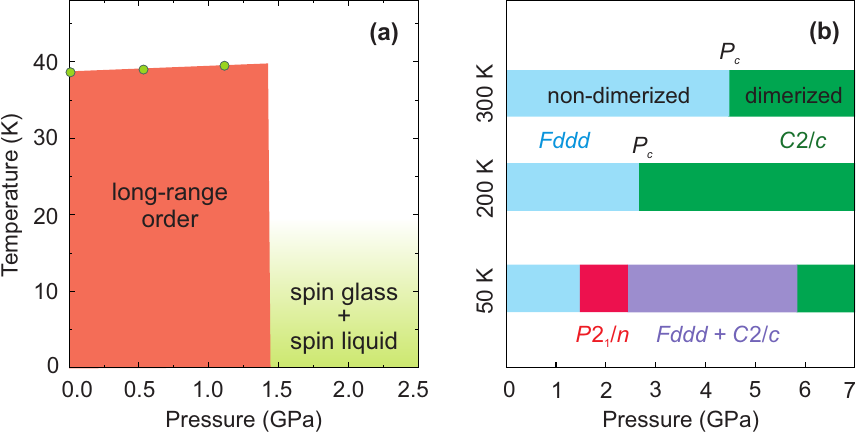}}
\caption{\label{fig:pressure}
Pressure evolution of $\beta$-Li$_2$IrO$_3$: (a) breakdown of magnetic order around 1.4\,GPa according to magnetic susceptibility and muon spin relaxation data~\cite{majumder2018}; (b) structural transformations from low-temperature x-ray diffraction~\cite{veiga2019}. Note that $P_c$ shifts toward lower pressures upon cooling. Additionally, an intermediate $P2_1/n$ phase and the phase mixture appear at low temperatures.
}
\end{figure}

Pressure evolution before the structural phase transition is peculiar too. Branching ratio measured by x-ray absorption reveals strong changes with pressure even below $P_c$~\cite{clancy2018,veiga2017} and suggests that the electronic state of Ir$^{4+}$, as well as magnetic interactions, may be altered significantly. In the case of $\alpha$-Li$_2$IrO$_3$, abrupt changes already below 0.2\,GPa were speculated based on powder data~\cite{clancy2018}, although concurrent and subsequent x-ray diffraction~\cite{hermann2018}, optical~\cite{hermann2019}, and Raman~\cite{li2020a} studies on single crystals did not reveal any additional structural phase transition below $P_c$. It is likely that $\alpha$-Li$_2$IrO$_3$ evolves non-monotonically under pressure, but retains its crystal structure till the onset of magnetism collapse and structural dimerization at $P_c$. Whether the incommensurately ordered magnetic ground state of $\alpha$-Li$_2$IrO$_3$ persists up to $P_c$ remains to be studied.

\begin{table*}
\caption{\label{tab:compounds}
Comparison of parent ($\alpha$-Li$_2$IrO$_3$, Na$_2$IrO$_3$) and substituted compounds: ionic radius $r$ for the six-fold coordinated ion~\cite{shannon1969}, lattice parameters ($a$, $b$, $c$, $\beta$), and unit cell volume $V$ per formula unit. Lattice parameters of Cu$_2$IrO$_3$ ($C2/c$) and $\alpha$-AIrO$_3$ with A = Mg, Zn, Cd ($R\bar 3$) have been re-calculated for the $C2/m$ unit cell of $\alpha$-Li$_2$IrO$_3$ to facilitate the comparison.
}
\centerline{%
\begin{tabular}{c@{\hspace{1cm}}c@{\hspace{1cm}}c@{\hspace{0.3cm}}c@{\hspace{0.3cm}}c@{\hspace{0.3cm}}c@{\hspace{0.6cm}}c@{\hspace{1cm}}c}\hline
    & $r$ (\r A) & $a$ (\r A)  & $b$ (\r A) & $c$ (\r A) & $\beta$ (deg) & $V$ (\r A$^3$) & Ref. \\\hline
 $\alpha$-Li$_2$IrO$_3$ & 0.68 & 5.175 & 8.936 & 5.119 & 109.83 & 55.68 & \cite{freund2016} \\
 Na$_2$IrO$_3$          & 0.97 & 5.427 & 9.395 & 5.614 & 109.04 & 67.65 & \cite{choi2012} \\\hline
 $\alpha$-MgIrO$_3$     & 0.66 & 5.158 & 8.935 & 4.418 & 90     & 53.46 & \cite{haraguchi2018} \\
 $\alpha$-ZnIrO$_3$     & 0.74 & 5.199 & 9.005 & 4.445 & 90     & 54.20 & \cite{haraguchi2018} \\
 $\alpha$-CdIrO$_3$     & 0.97 & 5.368 & 9.297 & 4.781 & 90     & 61.60 & \cite{haraguchi2020} \\
 Cu$_3$LiIr$_2$O$_6$    & 0.96 & 5.291 & 9.146 & 6.011 & 107.20 & 69.44 & \cite{roudebush2016} \\
 Cu$_3$NaIr$_2$O$_6$    & 0.96 & 5.390 & 9.300 & 5.971 & 108.64 & 71.28 & \cite{roudebush2016} \\
 Cu$_2$IrO$_3$          & 0.96 & 5.393 & 9.311 & 5.961 & 107.51 & 71.37 & \cite{abramchuk2017} \\
 Ag$_3$LiIr$_2$O$_6$    & 1.26 & 5.283 & 9.132 & 6.482 & 105.68 & 75.26 & \cite{bahrami2019} \\\hline
 H$_3$LiIr$_2$O$_6$     & --   & 5.349 & 9.243 & 4.873 & 111.44 & 56.07 & \cite{bette2017} \\\hline
\end{tabular}
}
\end{table*}

A different behavior was reported for the $\beta$-polymorph. Here, thermodynamic measurements and muon spin relaxation suggest the breakdown of magnetic order already at 1.4\,GPa, well below $P_c$, followed by the formation of a mixed state that blends static and dynamic spins in the form of coexisting spin glass and spin liquid (Fig.~\ref{fig:pressure}a)~\cite{majumder2018}. Interestingly, the dimerization pressure $P_c$ shows a strong temperature dependence~\cite{veiga2019}, but it does not seem that $P_c$ simply drops to 1.4\,GPa at low temperatures, making the breakdown of magnetic order a mere consequence of structural dimerization. More likely, a mixture of structurally dimerized and non-dimerized phases forms above 1.4\,GPa~\cite{veiga2019} and potentially explains the mixed magnetic state detected by muons~\cite{majumder2018}. Alternatively, a previously unforeseen intermediate phase with a more complex crystal structure may set in at 1.4\,GPa and precede the full dimerization (Fig.~\ref{fig:pressure}b)~\cite{veiga2019}. 

The behavior of $\beta$-Li$_2$IrO$_3$ right above 1.4\,GPa is thus of significant interest. It is so far the only Kitaev magnet where pressure was instrumental in stabilizing a magnetically disordered state without a complete loss of local magnetic moments. A similar breakdown of magnetic order may occur in $\gamma$-Li$_2$IrO$_3$ at a very similar pressure~\cite{breznay2017}, but its interplay with structural changes remains to be studied. 

Another important aspect of pressure tuning is that the reduction in the unit cell volume lowers the Ir--O--Ir angles and brings them closer to $90^{\circ}$ (Fig.~\ref{fig:geometry})~\cite{majumder2018}. Indeed, \textit{ab initio} calculations based on the experimentally determined structural parameters of $\beta$-Li$_2$IrO$_3$ under pressure revealed the enhancement of $\Gamma$ and the reduction in $|K|$~\cite{majumder2018}, the direction opposite to the one initially desired. Along with the problem of competing structural instabilities and imminent magnetism collapse, this renders hydrostatic pressure a less than optimal tuning strategy for reaching Kitaev limit in iridates. In fact, negative pressure or tensile strain~\cite{yadav2018} may be more promising in this case. 

Strain tuning has been discussed as a viable route toward the Kitaev limit also for $\alpha$-RuCl$_3$~\cite{kaib2021}, but experimental work in this direction remains scarce. Given that large strain would be required to reach the Kitaev limit, epitaxial growth on suitably chosen substrates should be the most promising strategy. Thin films of $\alpha$-Li$_2$IrO$_3$ have been grown on Y-doped ZrO$_2$ by pulsed laser deposition and proved to be polycrystalline in nature~\cite{jenderka2015}. In contrast, metal-organic aerosol deposition allows epitaxial growth on different substrates, although such films show mostly paramagnetic response down to low temperatures \cite{uhl2021}. Growing structurally ordered thin films of Kitaev iridates remains an interesting problem and will be a natural avenue for future research.


\subsection{Chemical substitutions}
\label{sec:substitutions}

\begin{figure*}
\centerline{\includegraphics[scale=0.9]{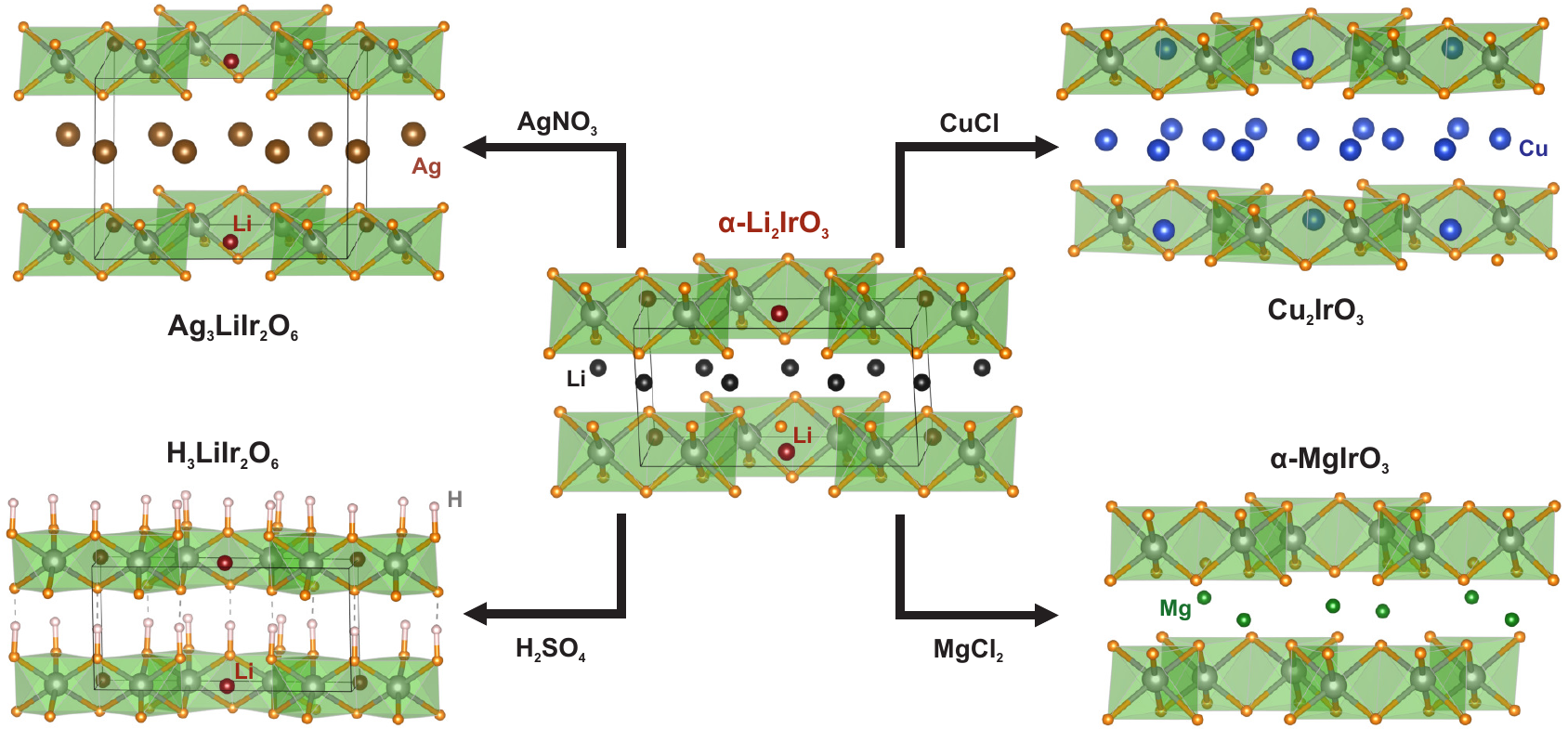}}
\caption{\label{fig:substitutions}
Chemical substitutions in $\alpha$-Li$_2$IrO$_3$. Parent structure contains Li atoms between the honeycomb layers (black), as well as in the centers of Ir hexagons (dark red). Only the former Li atoms are exchanged in Ag$_3$LiIr$_2$O$_6$ and H$_3$LiIr$_2$O$_6$, whereas a complete substitution is achieved in $\alpha$-MgIrO$_3$ and Cu$_2$IrO$_3$.
}
\end{figure*}

Pressure and strain effect can be also emulated by suitable chemical substitutions. Ion-exchange reactions have been used to prepare a gamut of new compounds where Li in $\alpha$-Li$_2$IrO$_3$ (or sometimes Na in Na$_2$IrO$_3$) is fully or partially replaced by other monovalent or divalent ions, see Table~\ref{tab:compounds} and Fig.~\ref{fig:substitutions}. The larger ions like Ag$^+$ may lead to only a partial (75\,\%) substitution because Li atoms centering Ir hexagons can not be exchanged. Other ions allow the complete substitution. All these compounds, except H$_3$LiIr$_2$O$_6$, are prepared by topotactic ion exchange at $300-400$\,$^{\circ}$C, well below the synthesis temperature of Li$_2$IrO$_3$. They are available as polycrystalline samples only. Similar to other compounds prepared by the ion exchange~\cite{tsirlin2012a,tsirlin2012b}, chemically substituted iridates are most likely metastable and should decompose if annealed at high temperatures.

Unit cell volumes of these compounds follow ionic radii of the substituted elements (Table~\ref{tab:compounds}). Smaller divalent metals (Mg, Zn) do not change the volume significantly. In contrast, monovalent metals lead to a significant expansion of the Li$_2$IrO$_3$ lattice. This further correlates with the presence of long-range magnetic order in $\alpha$-MgIrO$_3$ and $\alpha$-ZnIrO$_3$ below 32 and 47\,K, respectively~\cite{haraguchi2018}. On the other hand, most of the Cu- and Ag-substituted compounds show disordered magnetism of some sort~\cite{abramchuk2017,kenney2019,choi2019b,bahrami2019}, albeit with one exception~\cite{bahrami2021}.

At first glance, the suppression of long-range magnetic order in the Cu- and Ag-substituted iridates is consistent with the trend of approaching Kitaev limit upon lattice expansion. However, Na$_2$IrO$_3$ serves as another example of a similarly expanded $\alpha$-Li$_2$IrO$_3$ structure (Table~\ref{tab:compounds}), and it does show long-range magnetic order~\cite{singh2010,liu2011,ye2012}. In that case, nearest-neighbor $J$ and $\Gamma$ are indeed small\footnote{Note however that sizable $J$ and $\Gamma$ are used in Ref.~\cite{kim2020} to model the excitation spectra.} while $K$ is large and ferromagnetic in agreement with the preceding discussion, but third-neighbor interaction $J_3$ intervenes to stabilize antiferromagnetic zigzag order~\cite{winter2016}. Whether or not this $J_3$ appears also in the Cu- and Ag-substituted iridates remains an interesting problem for further investigation. In the case of the (Na$_{1-x}$Li$_x$)$_2$IrO$_3$ solid solutions that allow a gradual compression of Na$_2$IrO$_3$ up to $x\simeq 0.24$, long-range magnetic order persists~\cite{manni2014} suggesting the continuous presence of $J_3$. However, the type of magnetic order in Li-doped Na$_2$IrO$_3$ is not known with certainty~\cite{rolfs2015,simutis2018}. The miscibility gap above $x\simeq 0.24$ is caused by the fact that Li can be introduced in the honeycomb planes only (Fig.~\ref{fig:substitutions})~\cite{manni2014}.

Additionally, all substituted iridates are prone to stacking disorder~\cite{bette2019,kenney2019}. The $\alpha$-Li$_2$IrO$_3$ precursor is likely to contain stacking faults itself~\cite{manni,freund2016}, and their concentration can only increase when one removes ions that hold together the Ir honeycomb layers. Since the ion exchange takes place at a relatively low temperature, the system should be far from thermodynamic equilibrium and retain all the defects, both contained in the parent material and accumulated during the ion exchange. The problem of stacking disorder appears most prominently in Ag$_3$LiIr$_2$O$_6$ that has been prepared with different amounts of stacking faults. Samples with a low concentration of stacking faults show long-range magnetic order~\cite{bahrami2021,wang2020,chakraborty2021}, whereas samples with a high concentration of stacking faults develop a disordered magnetic state resembling a spin liquid~\cite{bahrami2019}. This observation is in line with the previous work on pure $\alpha$-Li$_2$IrO$_3$ where magnetic ordering transition is also suppressed when the concentration of stacking faults increases~\cite{manni}.

Yet another problem, now specific to the Cu substitution, is partial charge transfer. Experiments on Cu$_2$IrO$_3$ revealed that about 10\% of Cu$^+$ is oxidized into Cu$^{2+}$ with the respective amount of Ir$^{4+}$ reduced toward Ir$^{3+}$~\cite{kenney2019}. This brings one additional dimension, chemical disorder, which becomes even more prominent in hydrogenated lithium iridate, H$_3$LiIr$_2$O$_6$.


\subsection{Chemical disorder}

Cu$_2$IrO$_3$ features a mixture of four different species: Ir$^{4+}$, Cu$^{2+}$, Ir$^{3+}$, and Cu$^+$. The first two are magnetic, while the second two are not. It was argued that Cu$^{2+}$ resides in the center of Ir hexagons~\cite{kenney2019} and turns one of the adjacent Ir atoms into Ir$^{3+}$, thus diluting the honeycomb spin lattice~\cite{choi2019b}. Muons detect 60\% static and 40\% dynamic spins at low temperatures~\cite{kenney2019}. An enthusiastic interpretation of this behavior is that Cu$_2$IrO$_3$ without charge transfer would manifest the Kitaev spin liquid, but Cu$^{2+}$ ions create impurity centers where static spins precipitate~\cite{kenney2019}. On the other hand, in Ag$_3$LiIr$_2$O$_6$ a similar phenomenology is caused by stacking faults alone~\cite{bahrami2019,bahrami2021}, suggesting that structural disorder should play a crucial role in Cu$_2$IrO$_3$, too, and may be the only cause of its purported spin-liquid behavior. This abundant disorder will certainly complicate an unambiguous interpretation of continuum-like spectral features reported for Cu$_2$IrO$_3$ recently~\cite{takahashi2019,pal2020}.

Unlike other chemically substituted iridates, H$_3$LiIr$_2$O$_6$ is prepared hydrothermally by treating $\alpha$-Li$_2$IrO$_3$ in sulfuric acid~\cite{kitagawa2018}, and single crystals that even exceed $\alpha$-Li$_2$IrO$_3$ in size can be obtained~\cite{pei2020}. The compound lacks magnetic order, reveals persistent spin dynamics down to at least 50\,mK, and overall resembles a spin liquid~\cite{kitagawa2018}. On the other hand, not only are  stacking faults abundant~\cite{bette2017}, but also hydrogen atoms can take different positions between the Ir layers~\cite{li2018}. 

In contrast to monovalent metals, H$^+$ does not stay at the midpoint between the two layers and forms a short covalent bond to oxygen instead (Fig.~\ref{fig:substitutions}). This creates local O--H dipoles probed by dielectric measurements~\cite{wang2020b,geirhos2020}. At high temperatures, the H$^+$ ions tunnel between two equivalent sites in a paraelectric state. Below 100\,K, the slowing down of this proton dynamics~\cite{geirhos2020} leads to [LiIr$_2$O$_6$] honeycomb layers randomly decorated by hydrogen atoms in an effectively static configuration. \textit{Ab initio} calculations pinpoint strong changes in magnetic interactions depending on the positions of hydrogen and their bonding to oxygen~\cite{li2018,yadav2018b}, an effect also well-known for $3d$ transition metals~\cite{lebernegg2013}. 

Random positions of hydrogen should thus lead to an extreme randomness of magnetic interactions in H$_3$LiIr$_2$O$_6$. This effect alone is sufficient to break down long-range magnetic order and induce a magnetically disordered state not necessarily related to the Kitaev spin liquid. Abundant stacking faults~\cite{bette2017} will further facilitate this type of behavior. A prominent feature of H$_3$LiIr$_2$O$_6$ is a pileup of low-energy states probed by specific heat measurements and showing a strong dependence on the magnetic field. These states can be traced back to Majorana fermions in a bond-disordered Kitaev spin liquid~\cite{knolle2019}, although such phenomenology is by no means specific to this scenario~\cite{slagle2018}. It can be also accounted for by a random-singlet state arising from quenched disorder even in the absence of Kitaev interactions~\cite{kimchi2018}. Thermodynamic properties of the recently prepared Sr$_{0.5}$IrO$_3$ are well captured by this model~\cite{song2021}.

It is also instructive to review several studies that introduced disorder deliberately. Both magnetic Ru$^{4+}$~\cite{mehlawat2015} and non-magnetic Ti$^{4+}$~\cite{manni2014b} doping lead to a rapid suppression of the long-range order and formation of a glassy state. These observations further illustrate that even minor structural disorder may result in the absence of a magnetic transition. For honeycomb Na$_2$(Ir$_{1-x}$Ti$_x$)O$_3$, the spin-glass freezing temperature as well as the Curie-Weiss temperature extrapolate to zero at the site-percolation threshold of 30\%~\cite{manni2014b}, which is indeed expected in a honeycomb magnet with predominant nearest-neighbor couplings~\cite{andrade2014} and justifies the application of Eq.~\eqref{eq:jkg} in the microscopic analysis. On the other hand, for $\alpha$-Li$_2$(Ir$_{1-x}$Ti$_x$)O$_3$ the Curie-Weiss temperature as well as the freezing temperature clearly extend beyond this 30\% threshold up to at least $x\simeq 0.5$, thus giving evidence for interactions beyond nearest neighbors, which are indeed discussed theoretically for this compound~\cite{winter2016}. 

\section{Conclusions and outlook}

As the first decade of Li$_2$IrO$_3$ research comes to an end, it is time to summarize several important insights and lessons gained from this material, as well as outline possible directions for future studies:

\textit{First}, structural diversity of Li$_2$IrO$_3$ polymorphs has been the main trigger for exploring Kitaev model beyond planar honeycomb geometry. 3D versions of this model received broad attention only after they materialized in $\beta$ and $\gamma$-polymorphs of Li$_2$IrO$_3$. One exciting prediction of subsequent theoretical studies is the thermodynamic phase transition separating Kitaev spin liquid on the hyperhoneycomb lattice from the paramagnetic phase~\cite{nasu2014a,nasu2014b}. Experimental verification of this prediction pends a 3D hyperhoneycomb material, which is suitably tuned to the Kitaev limit.

\textit{Second}, Li$_2$IrO$_3$ polymorphs uncovered highly non-trivial magnetically ordered states with counter-rotating spin spirals. This incommensurate order is driven by anisotropic exchange terms and especially by the off-diagonal anisotropy $\Gamma$ that was previously not on radar. First experimental studies reveal magnons as excitations of these unusual incommensurate states at low energies, but also detect continuum-like features at higher energies where fractionalizion or magnon breakdown probably become relevant (Sec.~\ref{sec:excitations}). Detailed study of spectral properties for different Li$_2$IrO$_3$ polymorphs is clearly warranted, also from the perspective of topological magnons that are already predicted~\cite{choi2019c} as excitations of commensurate states obtained by quenching counter-rotating spirals in the applied field.

\textit{Third}, both 3D polymorphs of Li$_2$IrO$_3$ revealed a peculiar response to the applied field. They deviate from the intuitive behavior of Heisenberg antiferromagnets, which remain long-range-ordered in a spin-flop phase until fully saturated. In Kitaev magnets, exchange anisotropy and especially the off-diagonal $\Gamma$ term can lead to a quick suppression of magnetic long-range order in external field well before saturation is reached. This may create an impression that the material turns into a spin-liquid state, but experiments on $\beta$-Li$_2$IrO$_3$ clearly challenge such an interpretation. The field-induced state of this compound is a quantum paramagnet that does not break any symmetries and does not undergo any phase transitions as a function of temperature, but the underlying spin-spin correlations resemble conventional long-range order. This type of behavior can be easily confused with a field-induced spin-liquid state and requires caution in analyzing basic experimental signatures, such as presence or absence of thermodynamic phase transitions, in anisotropic magnets.

\textit{Fourth}, layered crystal structure of $\alpha$-Li$_2$O$_3$ has turned into a toolbox of new materials prepared by chemical substitution. It is probably not coincidental that these substitutions are easiest to perform in the material most prone to stacking faults and other types of disorder. Defects of the $\alpha$-Li$_2$IrO$_3$ precursor seem to be preserved and even enhanced upon chemical substitution. Several new Kitaev materials with the disordered magnetic ground state have been discovered, but randomness caused by structural disorder plays a crucial role in all of them. This structural disorder makes connections to the Kitaev spin liquid elusive even if these connections exist.

Ionic substitutions in 3D Li$_2$IrO$_3$ polymorphs could be a viable strategy, because those polymorphs are structurally well ordered and naturally immune to the stacking faults. Nevertheless, other types of defects should be carefully watched out. Many recent claims of the spin-liquid behavior are related to materials with structural imperfections of some sort, and instances of spin-liquid mimicry are not uncommon~\cite{zhu2017}. Spin dynamics driven by randomness can be interesting in its own right~\cite{li2020}, but manifold nature of defects in substituted iridates (stacking faults, Cu$^{2+}$ impurities, random positions of hydrogen) requires a suitable gauge to quantify the amount of disorder and its influence on exchange couplings. Strain tuning in epitaxial thin films and uniaxially compressed bulk crystals could be an alternative that eliminates the disorder or at least makes it easier to control. This interesting direction awaits further experimental development.

Li$_2$IrO$_3$ has been a challenging material for magnetism studies. Its central aspects, including magnetic order and field-induced transitions, could never be resolved using polycrystalline samples. Despite long and arduous efforts, typical crystal size remains well below 1\,mm for any of the polymorphs, but even these tiny crystals offered a comprehensive insight and firmly established the role of Kitaev interactions in this material. Several state-of-the-art techniques -- cantilever magnetometry, ac-calorimetry, resonant x-ray scattering (elastic and inelastic) -- have been instrumental in studying sub-mm size Li$_2$IrO$_3$ crystals. Further development of these techniques, along with continuous efforts in synthesis and crystal growth, will open new opportunities in understanding Li$_2$IrO$_3$ and its peculiar physics.
\medskip

\textbf{Acknowledgments} 

Fruitful research on Li$_2$IrO$_3$ and its derivatives would not be possible without continuous contributions and support from many of our colleagues, in particular Anton Jesche, Yogesh Singh, Friedrich Freund, Soham Manni, Alexander Zubtsovskii, Ina Pietsch, Mayukh Majumder, Rudra Manna, Radu Coldea, Stephen Winter, Roser Valent{\'\i}, and Ioannis Rousochatzakis. We acknowledge financial support by the German Research Foundation (DFG) via the Project No. 107745057 (TRR80). AAT was also supported by the Federal Ministry for Education and Research through the Sofja Kovalevskaya Award of Alexander von Humboldt Foundation.

%

\end{document}